\documentclass[journal]{IEEEtran}
\usepackage{graphicx}
\usepackage{amsmath}
\usepackage{amssymb} 
\usepackage{bm}
\usepackage[american]{circuitikz}
\usepackage{tikz}
\usepackage{tikz-3dplot}
\usepackage{subcaption}
\usepackage{caption}

\usetikzlibrary{babel, arrows.meta} 
\usepackage[utf8]{inputenc} 
\usepackage[utf8]{inputenc}
\usepackage{hyperref}
\usepackage{cuted}  
\usepackage{float}
\usepackage{tabularx}
\usepackage[]{hyperref}
\usepackage{multirow}
\usepackage{adjustbox}
\usepackage{xcolor} 
\usetikzlibrary{arrows.meta,decorations.markings,calc,shadows.blur}

\usepackage{xcolor}

\usetikzlibrary{arrows.meta,decorations.markings,calc,shadows.blur}
\title{Beyond Phasors: Solving Non-Sinusoidal Electrical Circuits using Geometry}

\author{Javier Castillo-Martínez, Raul Baños and  Francisco G. Montoya}

\begin{document}

\maketitle

\begin{abstract}

Classical phasor analysis is fundamentally limited to sinusoidal single-frequency conditions, which poses challenges when working in the presence of harmonics. Furthermore, the conventional solution, which consists of decomposing signals using Fourier series and applying superposition, is a fragmented process that does not provide a unified solution in the frequency domain. This paper overcomes this limitation by introducing a complete and direct approach for multi-harmonic AC circuits using Geometric Algebra (GA). In this way, all non-sinusoidal voltage and current waveforms are represented as simple vectors in a $2N$-dimensional Euclidean space. The relationship between these vectors is characterized by a single and unified geometric transformation termed the \textit{rotoflex}. This operator elevates the concept of impedance from a set of complex numbers per frequency to a single multivector that holistically captures the circuit response, while unifying the magnitude scale (flextance) and phase rotation (rotance) across all harmonics. Thus, this work establishes GA as a structurally unified and efficient alternative to phasor analysis, providing a more rigorous foundation for electrical circuit analysis. The methodology is validated through case studies that demonstrate perfect numerical consistency with traditional methods and superior performance.

\end{abstract}

\section{Introduction}
Complex numbers are widely used in electrical circuit analysis, a practice that originates from the phasor method introduced by Charles Proteus Steinmetz in 1893 \cite{steinmetz1893complex}.
Since then, its application has become widespread, significantly simplifying the analysis of sinusoidal alternating current (AC) circuits and networks under steady-state conditions \cite{kariyawasam2023assessment}. However, while early electrical systems were predominantly linear, for which the phasor method was entirely sufficient, the proliferation of non-linear loads is now ongoing and widespread. This trend is attributed to several key factors, including the widespread adoption of energy-efficient LED lighting \cite{montoya2016power} and the widespread use of power electronics in both household appliances and HVAC/R systems \cite{nunez2025harmonic}. Furthermore, a substantial contributor to this phenomenon is the industrial sector, where non-linear high-power loads, such as variable frequency drives (VFDs), electric arc furnaces, welding equipment, and electrolysis processes, are prevalent and significantly distort current waveforms \cite{das2002power}. Critically, the large-scale integration of renewable energy sources via their associated inverters and converters further aggravates the introduction of non-linearities in voltage and current waveforms \cite{blaabjerg2006overview}.

Within this operational context, the classical phasor method exhibits fundamental limitations, as its applicability is restricted to linear circuits with purely sinusoidal sources, and its analytical power is limited to calculations in the frequency domain at a single fundamental frequency \cite{alexander2021fundamentals}. The conventional workaround for non-sinusoidal signals involves decomposing the waveforms using the Fourier series and then applying superposition to analyze each harmonic individually \cite{carpinelli1996modeling}. This decomposition, however, exposes a core dimensional restriction of complex algebra: each harmonic must be represented in its own two-dimensional complex plane. These orthogonal frequency planes cannot be aggregated in the frequency domain using conventional complex numbers, which are inherently limited to represent planar quantities \cite{castro2019theorems}, forcing the final results to be reconstructed in the time domain \cite{menti2007geometric}. This constraint precludes a unified frequency domain solution and prevents direct corroboration of principles such as energy conservation across all harmonics \cite{castro2016m}. Furthermore, the traditional formalism offers limited geometric insight, as the imaginary unit $j$ is an abstract algebraic tool with no direct geometric equivalent. This fragments physical quantities such as power into separate real and imaginary parts, obscuring their nature as unified geometric entities. Several alternatives for analyzing non-linear systems exist, including simulations in the time domain, where the circuit is modeled as a system of non-linear integro-differential equations, which are then solved numerically step by step \cite{li2020review}. Another approach is state-space analysis, which remains relevant today in fields such as renewable energy grids \cite{han2021discrete}. Other methods, such as the dynamic phasor approach, have been proposed to unify the time and frequency domains but face significant limitations, including the assumption of slow-varying dynamics and inaccuracies in capturing the dynamic response \cite{belikov2018uses}.

While complex algebra has been the cornerstone of frequency-domain analysis, Geometric Algebra (GA) presents a more general and unifying mathematical system that contains complex numbers as a subset. In this context, Menti et al. \cite{menti2007geometric} first applied GA to represent power under non-sinusoidal conditions, introducing the "power multivector". Subsequently, a Generalized Complex Geometric Algebra (GCGA) was established to offer a more detailed geometric interpretation of apparent power \cite{castilla2008clifford}, with research extending to analysis of power flow with harmonics \cite{montoya2019analysis}. Furthermore, GA has been shown to subsume and generalize classical tools such as the Clarke and Park transformations \cite{montoya2022formulating} and has been applied to solve contemporary engineering challenges \cite{salmeron2024application}. Beyond power systems, its applicability has also been explored in signal processing \cite{montoya2022geometricapplied}. This body of work establishes GA not only as an alternative, but as a more encompassing language that unifies disparate concepts into a single, coherent, and geometrically intuitive structure.

While the application of Geometric Algebra (GA) to electrical circuit analysis has been addressed in a growing body of literature, a comprehensive framework that offers a direct solution based solely on GA principles has yet to be presented. Existing approaches often integrate GA as a conceptual or partial tool rather than as a self-contained end-to-end methodology. Thus, this paper addresses this gap by presenting, for the first time, a complete methodology for the frequency-domain analysis of electrical circuits under non-sinusoidal conditions using exclusively Geometric Algebra. In contrast to the classical phasor method, which requires a sequential, harmonic-by-harmonic analysis, this framework provides a direct solution that simultaneously accounts for all frequency components. This novel framework can be extended to any field within electrical engineering beyond circuit analysis, including power systems, electrical machines, or high-voltage transmission lines, to mention a few.

The main contributions of this approach are summarized as follows:
\begin{itemize}
    \item It establishes a unified algebraic structure that resolves the complete non-sinusoidal problem in a single operation, avoiding the traditional decomposition required by the Fourier series and superposition principle.
    \item It transcends the two-dimensional limitation of complex numbers, which can only represent a single harmonic in the Argand plane, by employing a $2N$-dimensional Euclidean space. This framework concurrently solves for $N$ harmonics  and is computationally more efficient than the classical per-harmonic complex-number solution.

    \item It preserves the intuitive link between the frequency and time domains by establishing a direct correspondence between its multivector elements and the complete non-sinusoidal waveforms, thus generalizing the relationship between a single phasor and a single sine wave.
     
\end{itemize}

The remainder of this article is structured as follows: Section II describes the fundamentals of geometric algebra, as well as the new operators proposed to address the problem posed; Section III formally presents the proposed method; Section IV presents the experimental analysis and shows how the results obtained in different test circuits support the advantages of the new method; and Section V summarizes the findings and suggests the potential of this methodology for future exploration in other electrical engineering applications.

\section{Theoretical Background}
\label{sec:theoretical_background}

This section outlines the mathematical foundations of Euclidean Geometric Algebra. First, the key concepts of the algebra are presented, followed by the definitions of its most relevant operations and properties.

\subsection{Fundamentals of Euclidean Geometric Algebra}
\label{subsec:ga_fundamentals}

Geometric Algebra (GA), built on the ideas of Grassmann and Clifford, offers a unified and geometrically intuitive mathematical framework \cite{Hestenes1986, Doran2003}.  This work specifically uses the  Euclidean Geometric Algebra  $\mathcal{G}_n$, generated from an $n$-dimensional real vector space. This space is endowed with an inner product and an orthonormal basis of vectors $\{\bm{\sigma}_1, \bm{\sigma}_2, \dots, \bm{\sigma}_n\}$ that satisfy the relation $\bm{\sigma}_i^2 = \bm{\sigma}_i \cdot \bm{\sigma}_i = 1$.

The elements of this algebra are constructed from basis elements, or \textit{blades}, which are classified by their grade, or geometric dimensionality. The simplest element is a grade-0 blade, which is a scalar ($\alpha, \beta \in \mathbb{R}$). A grade-1 blade is a vector representing a 1D oriented entity. Higher-grade blades are formed through the outer product; for example, a grade-2 blade, or bivector $\bm{a} \wedge \bm{b}$, represents an oriented plane. This concept is generalized to a grade blade $j$ or a $j$-vector, which represents an oriented $j$-dimensional subspace. The most general element in $\mathcal{G}_n$ is the multivector, a linear combination of blades of different grades:
\begin{equation}
    \bm{M} = \sum_{j=0}^{n} \langle \bm{M} \rangle_j
\end{equation}
where $\langle \bm{M} \rangle_j$ is the grade-$j$ part of the multivector $\bm{M}$.

The two operations that define the structure of GA are the outer product and the geometric product.

\subsubsection*{Outer Product ($\wedge$)}
The outer product of two vectors $\bm{a}$ and $\bm{b}$ generates a bivector $\bm{a} \wedge \bm{b}$. Its magnitude corresponds to the area of the parallelogram spanned by the vectors, and its orientation defines the plane they contain. It is associative and anticommutative for vectors:
\begin{equation}
    \bm{a} \wedge \bm{b} = -\bm{b} \wedge \bm{a}
\end{equation}

This property implies $\bm{a} \wedge \bm{a} = 0$, which encodes the idea that two collinear vectors do not define an area.

\subsubsection*{Geometric Product}
This is the central operation of GA and is denoted by juxtaposition. The geometric product of two vectors $\bm{a}$ and $\bm{b}$ is defined as:
\begin{equation}
    \bm{ab} = \bm{a} \cdot \bm{b} + \bm{a} \wedge \bm{b}
\end{equation}

This product combines the symmetric inner product (scalar part, $\bm{a} \cdot \bm{b}$) and the antisymmetric outer product (bivector part, $\bm{a} \wedge \bm{b}$). It is associative and, crucially, invertible for non-null vectors, a property that allows for division by vectors in a manner analogous to division by complex numbers.

\subsection{Properties and Notation}
\label{subsec:operations_notation}

From the geometric product, several highly useful operations are derived \cite{Doran2003}. The most important for this work are the reverse, the inverse, and rotation.

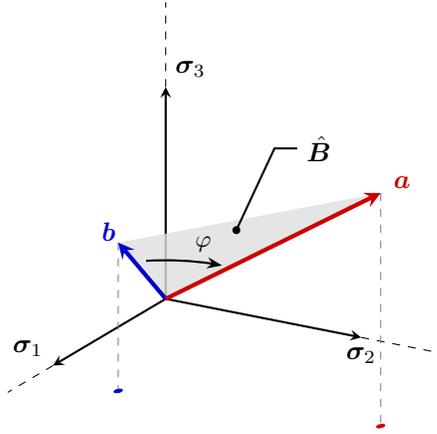
\begin{figure}[]
    \centering
    \tdplotsetmaincoords{70}{120}
    \begin{tikzpicture}[
        scale=3, 
        tdplot_main_coords,
        axis/.style={-stealth, thick, black},
        vector/.style={-stealth, ultra thick}
    ]
        \draw[axis] (0,0,0) -- (1,0,0) node[anchor=south east ]{$\bm{\sigma}_1$};
        \draw[axis] (0,0,0) -- (0,1,0) node[anchor=north ]{$\bm{\sigma}_2$};
        \draw[axis] (0,0,0) -- (0,0,1) node[anchor=south west]{$\bm{\sigma}_3$};
        \draw[dashed, thin] (1,0,0) -- (1.4,0,0);
        \draw[dashed, thin] (0,1,0) -- (0,1.4,0);
        \draw[dashed, thin] (0,0,1) -- (0,0,1.4);
        
        \tdplotsetcoord{U}{1.4}{60}{20}
        \tdplotsetcoord{V}{2.2}{60}{60}
        
        \coordinate (Uproj) at (Uxy);
        \coordinate (Vproj) at (Vxy);
        
        \draw[dashed, gray, thin] (U) -- (Uproj);
        \draw[dashed, gray, thin] (V) -- (Vproj);
        
        \draw[fill=blue!80!black, draw=blue!80!black] (Uproj) circle[x radius=0.025, y radius=0.015];
        \draw[fill=red!80!black, draw=red!80!black] (Vproj) circle[x radius=0.025, y radius=0.015];
        
        \fill[gray!25, opacity=0.8] (0,0,0) -- (U) -- (V) -- cycle;
        
        \tdplotdrawarc[->, thick, black, >={Stealth[length=2mm]}]{(0,0,0)}{0.5}{220}{180}{anchor=south west}{\small $\varphi$};
        
        \draw[vector, color=blue!80!black] (0,0,0) -- (U) node[pos=1.2] {$\bm{b}$};
        \draw[vector, color=red!80!black] (0,0,0) -- (V) node[pos=1.1] {$\bm{a}$};
        
        \coordinate (P) at (0.5, 0.65, 0.6); 
        \fill[black] (P) circle (0.5pt);
        \draw[thick] (P) -- ++(65:0.4cm) -- ++(0.1cm,0) node[anchor=west, black] {$\hat{\bm{B}}$};
    \end{tikzpicture}
        \caption{Rotation of a vector $\bm{b}$ to a vector $\bm{a}$ by an angle $\varphi$ within a plane $\hat{\bm{B}}$.}
    \label{fig:rotor_rotation}
\end{figure}

\begin{itemize}
    \item Reverse: The reverse operation, denoted by $\dagger$, consists of reversing the order of all vector products within a multivector. Given a $j$-vector $\bm{A}_j$, its reverse is defined as $\bm{A}_j^\dagger = (-1)^{j(j-1)/2} \bm{A}_j$. In this framework, the traditional complex conjugate is subsumed as a special case of the reverse operation.
    
    \item Inverse: Any non-null vector $\bm{a}$ has a unique multiplicative inverse, given by:
    \begin{equation}
        \bm{a}^{-1} = \frac{\bm{a}}{\bm{a}^2} = \frac{\bm{a}}{|\bm{a}|^2}
    \end{equation}

    \item Vector Rotation: One of GA's most powerful features is its ability to perform rotations algebraically. Rotations are universally handled in any dimension using an operator called a rotor, denoted by $\bm{R}$. A vector $\bm{b}$ is rotated into a vector $\bm{a}$ (of the same magnitude) through the sandwich product:
    \begin{equation}
        \bm{a} = \bm{R}_* \bm{b} \bm{R}_*^\dagger
        \label{eq:sandwich_rotor}
    \end{equation}
    
    The rotor itself can be constructed from the plane of rotation and the desired angle. For a rotation with an angle $\varphi$ in the plane defined by the unit bivector $\hat{\bm{B}}$ (see Fig~\ref{fig:rotor_rotation}), the rotor is formed using the half-angle:
    \begin{equation}
        \bm{R}_* = \cos\left(\frac{\varphi}{2}\right) + \hat{\bm{B}}\sin\left(\frac{\varphi}{2}\right) = e^{\frac{\varphi}{2}\hat{\bm{B}}}
    \end{equation}
    

\end{itemize}

To ensure clarity, the notation used throughout this paper is summarized in Table \ref{tab:notation}.

\begin{table}[]
\centering
\caption{Notation used in this paper.}
\label{tab:notation}
\begin{tabular}{l p{6cm}}
\hline
\textbf{Symbol} & \textbf{Description} \\
\hline
$\alpha, \beta, \dots$ & Scalars (lowercase Greek letters) \\
$\bm{a}, \bm{b}, \dots$ & Vectors (lowercase bold Latin letters) \\
$\bm{B}, \bm{F}, \dots$ & Bivectors (uppercase bold Latin letters) \\
$\bm{M}, \bm{N}, \bm{R}, \dots$ & Multivectors (uppercase bold Latin letters) \\
$\bm{a}\bm{b}$ & Geometric product of $\bm{a}$ and $\bm{b}$ \\
$\bm{a} \cdot \bm{b}$ & Inner product \\
$\bm{a} \wedge \bm{b}$ & Outer product \\
$\bm{M}^\dagger$ & Reverse of the multivector $\bm{M}$ \\
$\bm{M}^{-1}$ & Inverse of the multivector $\bm{M}$ \\
$|\bm{M}|$ & Magnitude of the multivector $\bm{M}$ \\
$\langle \bm{M} \rangle_j$ & Projection of the multivector $\bm{M}$ onto the grade-$j$ subspace \\
\hline
\end{tabular}
\end{table}

\section{Proposed Methodology}

\subsection{The GA Framework for AC Circuit Analysis}

Formally, this representation establishes a direct isomorphism between a time-domain signal and a GA vector. A non-sinusoidal signal $u(t)$, expanded into its Fourier series up to the N-th harmonic, is given by:
\begin{equation}
\begin{aligned}
  u(t) &= U_{dc} + \sqrt{2}\sum_{h=1}^{N} U_h\cos(h\omega t - \alpha_h)\\
  &= U_{dc} + \sqrt{2}\sum_{h=1}^{N} \left( U_{h,c} \cos(h\omega t) + U_{h,s} \sin(h\omega t) \right)  
\end{aligned}
\label{eq:fourier_series_def_revised}
\end{equation}
where the coefficients $U_{h,c}=U_h\cos\alpha_h$ and $U_{h,s}=U_h\sin\alpha_h$ are the RMS values of the cosine and sine components of the $h$-th harmonic, respectively. Ignoring the DC component for AC analysis, this set of $2N$ coefficients is mapped to the components of a single vector $\bm{u}$ in a $2N$-dimensional Euclidean space \cite{montoya2019analysis}:
\begin{equation}
\begin{aligned}
    \bm{u} &= \sum \bm{u}_h=\sum_{h=1}^{N} (u_{2h-1}\bm{\sigma}_{2h-1} + u_{2h}\bm{\sigma}_{2h}) \\
    &= \sum_{h=1}^N U_he^{-\alpha_h\hat{\bm{B}}_h}\bm{\sigma}_{2h-1}
\end{aligned}
\label{eq:ga_vector_def_revised}
\end{equation}
where $u_{2h-1} = U_{h,c}$, $u_{2h} = U_{h,s}$ and $\hat{\bm{B}}_h = \bm{\sigma}_{2h-1}\wedge \bm{\sigma}_{2h}=\bm{\sigma}_{2h-1,2h}$. 
The basis vectors $\{\bm{\sigma}_j\}$ are constant, orthonormal vectors, analogous to standard Cartesian basis vectors. The entire time dependence of the signal is thus encoded in this static vector representation, which captures its complete harmonic content. 

The analysis of alternating current (AC) circuits with linear loads is traditionally performed using complex numbers. Although effective for single-frequency conditions, the validity of this method collapses in the presence of harmonics. The conventional workaround involves decomposing the waveform via a Fourier Series and applying superposition, an inefficient process that requires solving the circuit for each harmonic individually. Geometric Algebra (GA) provides a powerful mathematical framework to overcome this limitation.  Thus, voltage and current are represented as \textit{vectors}, $\bm{u}$ and $\bm{i}$, in a geometric space where their Fourier coefficients serve as coordinates. In this formulation, the relationship between voltage and current becomes a purely geometric transformation, where the output vector is a scaled and rotated version of the input vector.

The most general form of this transformation is expressed through the sandwich product, which is robust for rotations in any dimension. For a generic scaling factor $k$ and a half-angle rotor $\bm{R}_{\star}$, the dual transformations can be written as:
\begin{equation}
\label{eq:sandwich_transform_both}
    \bm{i} = k \bm{R}_{\star} \bm{u} \bm{R}_{\star}^{\dagger}, \qquad
    \bm{u} = k^{-1} \bm{R}_{\star}^{\dagger} \bm{i} \bm{R}_{\star}
\end{equation}
While universally applicable, this operation can be simplified into a more compact, direct transformation. By choosing a rotor that operates within the plane defined by $\bm{u}$ and $\bm{i}$, the sandwich product simplifies it to a direct geometric product with a full-angle rotor, $\bm{R} = \bm{R}_{\star}^2$.
For practical circuit analysis, this direct transformation is adapted to two primary \color{black} reciprocal forms. The first,  used for \textit{series circuits}, \color{black} calculates the current from the voltage:
\begin{equation}
    \bm{i} = k_s \bm{R}_s \bm{u}
    \label{eq:series_form} 
\end{equation}
Here, $k_s$ is a scalar named the series \textit{flextance}, defined as $k_s = \lVert\bm{i}\rVert/\lVert\bm{u}\rVert$  (with units of Siemens, S),  and $\bm{R}_s$ is the series \textit{rotance} that performs the rotation from $\bm{u}$ to $\bm{i}$. 

\color{black}

The second form, used for \textit{parallel circuits}, calculates the voltage from the current:
\begin{equation}
    \bm{u} = k_p \bm{R}_p \bm{i}
    \label{eq:parallel_form}
\end{equation}
Here, $k_p$ is the parallel \textit{flextance}, noting its reciprocal definition $k_p = \lVert\bm{u}\rVert/\lVert\bm{i}\rVert$  (with units of Ohms, $\Omega$)  and the rotor $\bm{R}_p$ is the parallel rotance that rotates $\bm{i}$ to $\bm{u}$.

The primary advantage of this formulation is the ability to consolidate scaling and rotation into a single unified operator. This combined operator is named \textit{rotoflex}, denoted by $\bm{\Theta}$, which encapsulates the product of a scaling factor $k$ and a rotation operator $\bm{R}$:
\begin{equation}
    \bm{\Theta} = k\bm{R}
\end{equation}

This operator admits two fundamental realizations depending on the circuit topology: the \textit{series rotoflex} $\bm{\Theta}_s = k_s\bm{R}_s$ yielding $\bm{i} = \bm{\Theta}_s \bm{u}$, and the \textit{parallel rotoflex} $\bm{\Theta}_p = k_p\bm{R}_p$ yielding $\bm{u} = \bm{\Theta}_p \bm{i}$. Since the rotance $\bm{R}$ is a dimensionless geometric operator, the units of the rotoflex $\bm{\Theta}$ are identical to those of its flextance $k$.  Both forms are entirely consistent and provide full analytical flexibility through their corresponding inverse relations:

\color{black}

\begin{equation}
    \bm{\Theta}^{-1} = (k\bm{R})^{-1} = k^{-1}\bm{R}^{-1} = k^{-1}\bm{R}^{\dagger}
    \label{eq:rotoflex_inverse}
\end{equation}

This inverse property enables bidirectional computation between voltage and current in both topologies: $\bm{u} = \bm{\Theta}_s^{-1} \bm{i}$ for the series case, and $\bm{i} = \bm{\Theta}_p^{-1} \bm{u}$ for the parallel case.

The primary contribution of this work, therefore, is to derive these specific rotoflex operators for fundamental $RLC$ loads. By doing so,  it lays the foundation for a paradigm that transcends the dimensional restriction of complex algebra, enabling for the first time a unified solution in the frequency domain. This approach relax the need to reconstruct the final response in the time domain, positioning Geometric Algebra as a conceptually superior framework compared to phasor analysis for the analysis of modern electrical circuits.

\subsection{Calculation of the Flextance}

The calculation of the scaling factors, or flextances, depends on two key components: the harmonic profile of the known input signal and the frequency response of the circuit's topology. To present this calculation in a unified framework, it is defined a generalized approach that applies to both series and parallel circuits.

The first component required is the \textit{spectral weight}, $\gamma_h$, which represents the normalized magnitude of each harmonic component $h$ of the known input signal signal ${x(t)}$. This signal in vector form will be the voltage $\bm{u}$ for the series case and the current $\bm{i}$ for the parallel case.
\begin{equation}
    \gamma^2_{h} = \frac{x_{2h-1}^2+x_{2h}^2}{\lVert\bm{x}\rVert^2} = \frac{x_{2h-1}^2+x_{2h}^2}{{\sum_{j=1}^{2N} x_j^2}} = \frac{X_h^2}{\sum_{j=1}^{2N} x_j^2}
    \label{eq:gamma_generic_definition}
\end{equation}

The second component is the \textit{spectral kernel}, $\kappa_h$, which encodes the circuit's frequency-dependent response to each harmonic. Its squared magnitude is defined generally based on the circuit's dissipative and energy storing properties at the $h$-th harmonic:
\begin{equation}
    \kappa_h^2 = \frac{1}{\mathcal{D}_h^2 + \mathcal{X}_h^2} 
    \label{eq:unified_kernel_def}
\end{equation}
where $\mathcal{D}_h$ represents the harmonic \textit{dissipance}, the dissipative component (related to resistance $R$ or conductance $G$), and $\mathcal{X}_h$ represents the \textit{storance}, the net harmonic energy storing component (resulting from the interplay between magnetic energy storage in inductors and electric energy storage in capacitors). The specific forms of $\mathcal{D}_h$ and $\mathcal{X}_h$ depend on the considered circuit topology, as detailed below.

With these generalized components, the squared flextance $k$ is computed as a weighted square of the spectral kernel values:
\begin{equation}
    k^2 = {\sum_{h=1}^{N} \gamma_{h}^2 \kappa_h^2}
    \label{eq:k_unified_calc}
\end{equation}
This unified formulation can now be specialized for series and parallel topologies.

\subsubsection{Case 1 - Series Circuit}
When a series circuit (Fig.~\ref{fig:case1_sub}) is driven by a known voltage vector $\bm{u}$, the flextance $k_s$ represents an effective admittance-like ($\lVert\bm{i}\rVert/\lVert\bm{u}\rVert$). The spectral weights, $\gamma_h$, are computed from the components of $\bm{u}$.

For this topology, the dissipance and storance components correspond to  $\mathcal{D}_h = R$ and $\mathcal{X}_h = h\omega L - \frac{1}{h\omega C}$, respectively.
Substituting these into Eq.~\eqref{eq:unified_kernel_def} yields the \textit{series spectral kernel}
\begin{equation}
    \kappa_{s,h}^2 = \frac{1}{R^2 + \left(h\omega L - \frac{1}{h\omega C}\right)^2}
    \label{eq:sh_kernel_def}
\end{equation}
The series flextance $k_s$ is then computed by applying the general formula from Eq.~\eqref{eq:k_unified_calc}:
\begin{equation}
    k_s = \sqrt{\sum_{h=1}^{N} \gamma_{h}^2 \kappa_{s,h}^2}
    \label{eq:ks_final_calc}
\end{equation}

\subsubsection{Case 2 - Parallel Circuit}
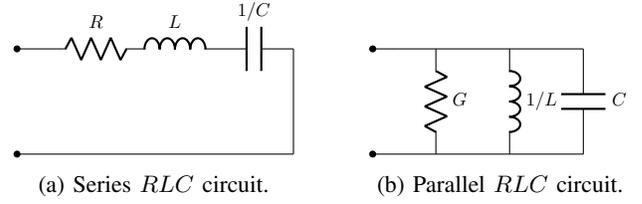
\begin{figure}[]
    \centering
    \begin{subfigure}[b]{0.48\columnwidth}
        \centering
        \begin{circuitikz}[scale=0.7, transform shape]
            \draw (0,0) node[circ] {} to (0.75,0);
            \draw (0,2) node[circ] {} to (0.75,2);
            \draw (0.75,2) to[R, l=$R$] (2.25,2) to[L, l=$L$] (3.75,2) to[C, l=$1/C$] (5.25,2);
            \draw (0.75,0) -- (5.25,0);
            \draw (5.25,2) -- (5.25,0);
        \end{circuitikz}
        \caption{Series $RLC$ circuit.}
        \label{fig:case1_sub}
    \end{subfigure}
    \hfill
    \begin{subfigure}[b]{0.48\columnwidth}
        \centering
        \begin{circuitikz}[scale=0.7, transform shape]
            \draw (0,0) node[circ] {} to (0.5,0);
            \draw (0,2) node[circ] {} to (0.5,2);
            \draw (0.5,2) -- (4,2); \draw (0.5,0) -- (4,0);
            \draw (1.2,2) to[R, l=$G$] (1.2,0);
            \draw (2.6,2) to[L, l=$1/L$] (2.6,0);
            \draw (4.0,2) to[C, l=$C$] (4.0,0);
        \end{circuitikz}
        \caption{Parallel $RLC$ circuit.}
        \label{fig:case2_sub}
    \end{subfigure}
    \caption{The two fundamental topologies under study.}
    \label{fig:rlc_topologies_combined}
\end{figure}

Conversely, when a parallel circuit (Fig.~\ref{fig:case2_sub}) is driven by a known current vector $\bm{i}$, the flextance $k_p$ represents an effective impedance-like component ($\lVert\bm{u}\rVert/\lVert\bm{i}\rVert$). The spectral weights, $\gamma_h$, are computed from the components of $\bm{i}$.

For this topology, the dissipative and energy storing components correspond to  $\mathcal{D}_h = G$ and $\mathcal{X}_h = h\omega C - \frac{1}{h\omega L}$.
Substituting these into the unified kernel definition yields the \textit{parallel spectral kernel}
\begin{equation}
    \kappa_{p,h}^2 = \frac{1}{G^2 + \left(h\omega C - \frac{1}{h\omega L}\right)^2}
    \label{eq:ph_kernel_def}
\end{equation}
The parallel flextance $k_p$ is computed accordingly:
\begin{equation}
    k_p = \sqrt{\sum_{h=1}^{N} \gamma_{h}^2 \kappa_{p,h}^2}
    \label{eq:kp_final_calc}
\end{equation}

\subsection{Calculation of the Rotance Operator}

The rotance operator, $\bm{R}$ is the rotational component of the corresponding rotoflex operator, $\bm{\Theta}$. Geometrically, the rotance performs the necessary rotation to match the orientation of the output vector with respect to the input. This is a single rotation in the $2N$-dimensional space, which is composed of independent rotations within each harmonic plane.

The calculation of this overall rotor requires knowing both the full input and output vectors. The procedure, therefore, begins by first constructing the unknown output vector. This is achieved by applying the known physical response of the circuit to each harmonic of the input vector individually. The $h$-th harmonic of a voltage vector, $\bm{u}_h$, and its corresponding current vector, $\bm{i}_h$, are vectors confined to the plane spanned by $\{\bm{\sigma}_{2h-1}, \bm{\sigma}_{2h}\}$  (see Eq.~\eqref{eq:ga_vector_def_revised}).
The specific construction of the unknown harmonic vector depends on the circuit topology, as detailed in the following cases.

\subsubsection{Case 1 - Series Circuit}
When a series circuit is driven by a known voltage vector $\bm{u}$, each harmonic component of the output current, $\bm{i}_h$, is obtained by scaling its corresponding voltage harmonic $\bm{u}_h$ by the spectral kernel magnitude $\kappa_{s,h}$ and rotating it by the harmonic rotor $\bm{R}_h$:
\begin{equation}
    \bm{i}_h = \kappa_{s,h} \bm{R}_{s,h} \bm{u}_h
\end{equation}
The magnitude and angle of rotation are determined by the circuit's physical parameters:
\begin{align*}
    \bm{R}_{s,h} &= e^{\varphi_{s,h} \hat{\bm{B}}_h}= e^{\varphi_{s,h} {\bm{\sigma}_{2h-1,2h}}} \\
   \kappa_{s,h} &= 1/\sqrt{R^2 + (h\omega L - 1/(h\omega C))^2} \\
   \varphi_{s,h} &= \arctan\left(\frac{1/(h\omega C)-h\omega L}{R}\right)
\end{align*}

\color{black}

This geometric operation is illustrated in Fig.~\ref{fig:harmonic_transformation}. The input voltage harmonic $\bm{u}_h$, with an initial phase $\alpha_h$, is rotated by the angle $\varphi_h$ introduced by the circuit's elements, resulting in the output current harmonic $\bm{i}_h$. By summing these constructed harmonic vectors, it is obtained the complete output current vector, $\bm{i} = \sum_h \bm{i}_h$. {Once the full input vector $\bm{u}$ and the full constructed output vector $\bm{i}$ are known}, the overall series rotance is computed directly from their normalized forms:
\begin{equation}
\bm{R}_s = \hat{\bm{\imath}}\hat{\bm{u}} \quad (\text{rotates } \bm{u} \text{ to } \bm{i})
\end{equation}

This direct rotor is related to the half-angle sandwich rotors ($\bm{R}_{\star}$) by the identity $\bm{R}_s = \bm{R}_{\star}^2$. Its well-known spinor form is given by $\bm{R}_s = \cos\varphi_s + \hat{\bm{B}}_s\sin\varphi_s$, where $\varphi_s$ is the effective angle between the two total vectors and $\hat{\bm{B}}_s$ is the unit bivector representing the overall plane of rotation for the series case.

\subsubsection{Case 2 - Parallel Circuit}
Conversely, for a parallel topology driven by a known current vector $\bm{i}$, each voltage harmonic $\bm{u}_h$ is constructed by scaling the current harmonic $\bm{i}_h$ and applying the rotation defined by the rotance $\bm{R}_{p,h}$:
\begin{equation}
 \bm{u}_{h} = \kappa_{p,h} \bm{R}_{p,h} \bm{i}_h
\end{equation}
where the parameters are now defined as:
\begin{align*}
 \bm{R}_{p,h} &= e^{\varphi_{p,h} \hat{\bm{B}}_h} \\
 \kappa_{p,h} &= 1/\sqrt{G^2 + (h\omega C - 1/(h\omega L))^2} \\
 \varphi_{p,h} &=\arctan \left(\frac{1/(h\omega L)-h\omega C}{G}\right)
\end{align*}

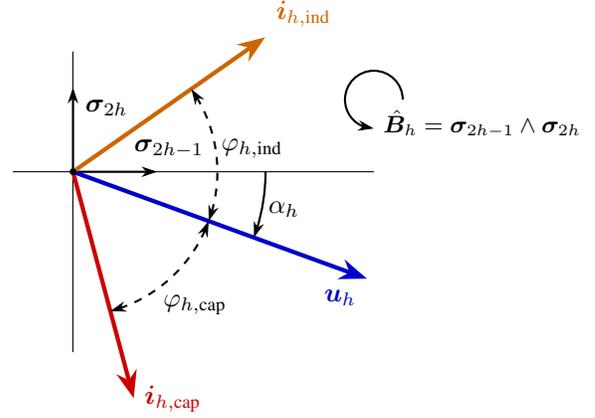
\begin{figure}[]
    \centering
    \begin{tikzpicture}[scale=1.6, >={Stealth[length=6pt, width=4pt]}]
        \draw[-] (-0.5,0) -- (2.5,0); 
        \draw[-] (0,-1.5) -- (0,1);
        \draw[->, thick] (0,0) -- (0.7,0); 
        \draw[->, thick] (0,0) -- (0,0.7) node[below right=0.1em] {$\bm{\sigma}_{2h}$};

        \def\angleA{20} 
        \def\angleP{55} 

        \draw[->, ultra thick, blue!80!black, scale=1.3, -{Stealth[scale=1.02]}] (0,0) -- (-\angleA:2.0) node[pos=1, anchor=north east] {$\bm{u}_h$};

        \draw[->, ultra thick, red!80!black,scale=1.3, -{Stealth[scale=1.02]}] (0,0) -- (-\angleA-\angleP:1.5) node[pos=1, anchor=west] {$\bm{i}_{h, \text{cap}}$};

        \draw[->, ultra thick, orange!80!black, scale=1.3, -{Stealth[scale=1.02]}] (0,0) -- (-\angleA+\angleP:1.5) node[pos=1, anchor=south west] {$\bm{i}_{h, \text{ind}}$};

        \draw[->, black, thick] (1.6,0) arc (0:-\angleA:1.6);
        \node[black] at (-\angleA/2:1.78) {$\alpha_h$}; 

        \draw[<->, black, thick, dashed] (-\angleA:1.2) arc (-\angleA:-\angleA-\angleP:1.2);
        \node[black] at (-\angleA-\angleP/2:1.5) {$\varphi_{h, \text{cap}}$};

        \draw[<->, black, thick, dashed] (-\angleA:1.2) arc (-\angleA:-\angleA+\angleP:1.2);
        \node[black] at (-\angleA+\angleP/2:1.5) {$\varphi_{h, \text{ind}}$};
        
        \begin{scope}[shift={(2.5,0.6)}, scale=0.2] 
            \draw[thick, <-] (270:1.2) arc (270:0:1.2);
            \node[anchor=center, font=\small] at (4.5,-1.0) {$\hat{\bm{B}}_h=\bm{\sigma}_{2h-1}\wedge\bm{\sigma}_{2h}$}; 
        \end{scope}

        \fill (0,0) circle (0.8pt); 

        \node[black, anchor=south] at (3:0.8) {$\bm{\sigma}_{2h-1}$};
    \end{tikzpicture}
    \caption{Harmonic transformation for a series circuit. The input voltage vector $\bm{u}_h$ is transformed into an output current vector. A capacitive circuit applies a clockwise rotation $\varphi_{h, \text{cap}}$, while an inductive circuit applies a counter-clockwise rotation $\varphi_{h, \text{ind}}$.}
    \label{fig:harmonic_transformation}
\end{figure}

This corresponds to an inverse rotation from the current vector to the voltage vector, as depicted conceptually in Fig.~\ref{fig:harmonic_transformation}. The complete output voltage vector is obtained by summing the constructed harmonic vectors, $\bm{u} = \sum_h \bm{u}_h$. With both the full input vector $\bm{i}$ and the complete output vector $\bm{u}$ defined, the overall parallel rotance is computed as:
\begin{equation}
\bm{R}_p = \hat{\bm{u}}\hat{\bm{\imath}} \quad (\text{rotates } \bm{i} \text{ to } \bm{u})
\end{equation}

This operator, like its series counterpart, represents the single geometric rotation in the multi-harmonic space that maps the input vector to the output vector for the parallel circuit topology. Its spinor form is $\bm{R}_p = \cos\varphi_p + \hat{\bm{B}}_p\sin\varphi_p$.

\section{Validation and Results}
\subsection{Baseline Validation: Single-Frequency Case}
\label{subsec:single_freq_validation}

Before demonstrating the framework's multi-harmonic capabilities, it is first proved  that it correctly reproduces classical phasor results in the degenerate (single-frequency) case ($N=1$). This establishes the GA methodology as a proper generalization of traditional analysis.

\subsubsection{Series RLC Circuit}
Consider a series RLC circuit excited by $u(t) = U\sqrt{2}\cos(\omega t)$, which maps to $\bm{u} = U\bm{\sigma}_1$. Applying the series rotoflex operator $\bm{\Theta}_s = k_s\bm{R}_s$ with $k_s = 1/\sqrt{R^2 + X_1^2}$ (where $X_1 = \omega L - 1/(\omega C)$) and $\bm{R}_s = k_s(R - X_1\bm{\sigma}_{12})$, it is obtained:
\begin{equation}
\bm{i} = \frac{UR}{R^2 + X_1^2}\bm{\sigma}_1 + \frac{UX_1}{R^2 + X_1^2}\bm{\sigma}_2
\end{equation}

This is exactly equivalent to the classical phasor solution 

$$\mathbf{I} = \frac{UR}{(R^2+X_1^2)} - j\frac{UX_1}{(R^2+X_1^2)}$$
under the mapping $\mathbf{X} = A-jB \longleftrightarrow \bm{x} = A\bm{\sigma}_1 + B\bm{\sigma}_2$.

\subsubsection{Parallel RLC Circuit}
Similarly, for a parallel RLC circuit with $i(t) = I\sqrt{2}\cos(\omega t)$ mapping to $\bm{i} = I\bm{\sigma}_1$, the parallel operator yields:
\begin{equation}
\bm{u} = \frac{IG}{G^2 + B_1^2}\bm{\sigma}_1 + \frac{IB_1}{G^2 + B_1^2}\bm{\sigma}_2
\end{equation}
where $B_1 = \omega C - 1/(\omega L)$, again matching the classical result perfectly.

These validations confirm that the GA framework contains traditional phasor analysis as a special case. Supplementary material for reproducing the results of this paper are available at 
https://git.new/RotoFlex. 

Having established this baseline equivalence, it is now demonstrated the main advantage of the proposed method, that is, a method for direct multi-harmonic circuit analysis without decomposition or superposition.

\subsection{Multi-Harmonic Case Studies}
\label{subsec:multi_harmonic_cases}

To demonstrate the framework's core capability—direct solution of multi-harmonic circuits—they are now presented two  representative case studies and compare the results against classical per-harmonic superposition. Beyond numerical validation, it is  analyzed the geometric interpretation of the resulting operators, highlighting the physical insight provided by the GA framework.

\subsubsection{Case 1: Series RLC Circuit with Two Harmonics}
\label{subsubsec:case1_series}

Consider a series RLC circuit with parameters $R=3\,\Omega$, $L=1\,\text{H}$, and $C=1\,\text{F}$, driven by a two-harmonic voltage source at fundamental frequency $\omega=1\,\text{rad/s}$:
\begin{equation}
u(t)=\sqrt{2}\cos(\omega t)+0.8\sqrt{2}\cos(2\omega t)\,\text{V}
\end{equation}
Both harmonics are in phase ($\alpha_1=\alpha_2=0$), mapping to the GA vector $\bm{u}=\bm{\sigma}_1+0.8\bm{\sigma}_3$ (components in Volts).

The classical solution requires separate analysis of each harmonic using complex impedances $\mathbf{Z}_h = R + j(h\omega L - 1/(h\omega C))$. For the fundamental, the purely resistive impedance $\mathbf{Z}_1 = 3\,\Omega$ yields $\mathbf{I}_1 = 0.3333\angle 0^\circ\,\text{A}$. The second harmonic encounters a reactive impedance $\mathbf{Z}_2 = 3 + j1.5\,\Omega$, resulting in $\mathbf{I}_2 = 0.2386\angle -26.57^\circ\,\text{A}$. These individual phasor results must then be reconstructed in the time domain or mapped to separate GA subspaces and summed.

In contrast, the GA framework computes the series flextance $k_s = 0.3201\,\text{S}$ using Eq.~\eqref{eq:ks_final_calc}, which represents the overall magnitude scaling across both harmonics. The series rotance $\bm{R}_s$ encodes the complete phase relationship between the total voltage and current waveforms. Its explicit form is:
\begin{equation}
\bm{R}_s = 0.9602 + 0.1016\bm{\sigma}_{13} - 0.2032\bm{\sigma}_{14} - 0.1626\bm{\sigma}_{34}
\label{eq:case1_rotance}
\end{equation}

The scalar component $\langle\bm{R}_s\rangle_0 = 0.9602$ represents the in-phase alignment between voltage and current and directly yields the power factor $\text{PF} = 0.9602$. The three bivector terms $\bm{\sigma}_{13}$, $\bm{\sigma}_{14}$, and $\bm{\sigma}_{34}$ define the composite rotation plane in the four-dimensional $\mathcal{G}_4$ space ($2N=4$ for $N=2$ harmonics). Geometrically, this corresponds to a rotation by an effective angle $\varphi_{\text{eff}} = \arccos(0.9602) = 16.26^\circ$ within a plane that simultaneously accounts for the phase shifts of both harmonics. The presence of multiple bivector components reflects the fact that the two harmonics interact with the circuit's frequency-dependent reactance differently, creating a composite geometric relationship that cannot be captured by a single complex number.

Applying the rotoflex operator $\bm{\Theta}_s = k_s\bm{R}_s$ to the voltage vector yields the current directly: $\bm{i} = 0.3333\bm{\sigma}_1+0.2133\bm{\sigma}_3+0.1067\bm{\sigma}_4$ (components in Amperes). Notably, the $\bm{\sigma}_4$ component emerges from the geometric interaction encoded in $\bm{R}_s$ and represents the out-of-phase contribution of the second harmonic. The GA solution is numerically identical to the classical superposition result, but achieves this through a single unified transformation rather than sequential per-harmonic calculations. 

\subsubsection{Case 2: Parallel RLC Circuit with Three Harmonics}
\label{subsubsec:case2_parallel}

The second case study examines a parallel RLC circuit with $R = 2\,\Omega$ ($G = 0.5\,\text{S}$), $L = 3\,\text{H}$, and $C = 0.5\,\text{F}$, excited by a three-harmonic current source at $\omega=2\,\text{rad/s}$:
\begin{equation*}
i(t) = 1.5\sqrt{2}\cos(\omega t) + 0.9\sqrt{2}\sin(2\omega t) + 0.5\sqrt{2}\cos(3\omega t)\,\text{A}
\end{equation*}
The presence of the sine term in the second harmonic introduces an additional phase shift of $-90^\circ$, increasing the geometric complexity. This maps to $\bm{i} = 1.5\bm{\sigma}_1 + 0.9\bm{\sigma}_4 + 0.5\bm{\sigma}_5$ (components in Amperes).

The classical approach again requires three separate calculations using per-harmonic admittances $\mathbf{Y}_h = G + j(h\omega C - 1/(h\omega L))$. The susceptance varies significantly across harmonics due to the frequency-dependent reactive terms, resulting in voltage phasors with different magnitudes and angles: $\mathbf{U}_1 = 0.7941 - j1.3235\,\text{V}$, $\mathbf{U}_2 = -0.4396 - j0.1147\,\text{V}$, and $\mathbf{U}_3 = 0.0280 - j0.1651\,\text{V}$. The final solution is obtained by summing these components across their respective frequency subspaces.

The GA framework computes the parallel flextance $k_p = 0.8892\,\Omega$ and the parallel rotance $\bm{R}_p$, which now operates in the six-dimensional space $\mathcal{G}_6$ ($2N=6$ for $N=3$ harmonics). The rotance operator contains 12 independent bivector components, reflecting the geometric complexity of the three-harmonic interaction. 

\begin{equation*}
\begin{split}
\bm{R}_p ={}& 0.4446 - 0.6746\bm{\sigma}_{12} + 0.2241\bm{\sigma}_{13} + 0.1844\bm{\sigma}_{14} \\
& + 0.1206\bm{\sigma}_{15} - 0.0841\bm{\sigma}_{16} + 0.4047\bm{\sigma}_{24} + 0.2249\bm{\sigma}_{25} \\
& - 0.1344\bm{\sigma}_{34} - 0.0747\bm{\sigma}_{35} + 0.0109\bm{\sigma}_{45} - 0.0505\bm{\sigma}_{46} \\
& - 0.0280\bm{\sigma}_{56}
\end{split}
\end{equation*}

The scalar component $\langle\bm{R}_p\rangle_0 = 0.4446$ indicates a significantly lower power factor of 0.4446, corresponding to an effective rotation angle of $\varphi_{\text{eff}} = 63.58^\circ$. This large angle reflects the substantial phase displacement introduced by the reactive circuit across the three frequency components.

The bivector structure of $\bm{R}_p$ encodes how each harmonic's phase relationship with the load combines to form the overall geometric transformation. For instance, the dominant term $-0.6746\bm{\sigma}_{12}$ captures the strong reactive behavior at the fundamental frequency, while smaller cross-terms like $\bm{\sigma}_{13}$ and $\bm{\sigma}_{45}$ represent the coupling between different harmonic subspaces. This geometric representation provides direct physical insight: the operator is not merely a list of complex numbers for each frequency, but a single geometric entity that describes the circuit's behavior as a unified multivector transformation.

The direct application $\bm{u} = k_p\bm{R}_p\bm{i}$ yields the full voltage vector directly: $\bm{u} = 0.7941\bm{\sigma}_{1} + 1.3235\bm{\sigma}_{2} - 0.4396\bm{\sigma}_{3} + 0.1147\bm{\sigma}_{4} + 0.0280\bm{\sigma}_{5} + 0.1651\bm{\sigma}_{6}$ (components in Volts). The solution is again numerically identical to the classical result, but the GA formulation reveals the underlying geometric structure that remains hidden in the traditional frequency-by-frequency approach. 

\subsubsection{Comparative Analysis}

Table~\ref{tab:comparison_cases} summarizes the numerical results for both case studies, comparing the classical phasor method against the proposed GA framework. The magnitude and phase of each harmonic component are provided for direct comparison, along with the overall power factor computed by each method. In both cases, the agreement is exact within numerical precision, validating the accuracy of the GA approach.

\begin{table}[]
\centering
\caption{Comparison of Classical Phasor and GA Results for Multi-Harmonic Case Studies}
\label{tab:comparison_cases}
\setlength{\tabcolsep}{4pt}
\renewcommand{\arraystretch}{1.3}
\begin{tabular}{ccccr@{.}l}
\hline
\textbf{Case} & \textbf{Harmonic} & \textbf{Method} & \textbf{Magnitude} & \multicolumn{2}{c}{\textbf{Phase ($^\circ$)}} \\
\hline
\multirow{6}{*}{\textbf{1 (Series)}} 
& \multirow{2}{*}{$h=1$} & Classical & 0.33 A & 0&00 \\
& & GA & 0.33 A & 0&00 \\
\cline{2-6}
& \multirow{2}{*}{$h=2$} & Classical & 0.24 A & $-26$&57 \\
& & GA & 0.24 A & $-26$&57 \\
\cline{2-6}
& \multicolumn{5}{c}{{Effective rotation (GA): $\varphi_{\text{eff}} = 16.26^\circ$}, PF = 0.96} \\
\hline
\multirow{8}{*}{\textbf{2 (Parallel)}} 
& \multirow{2}{*}{$h=1$} & Classical & 1.54 V & $-59$&04 \\
& & GA & 1.54 V & $-59$&04 \\
\cline{2-6}
& \multirow{2}{*}{$h=2$} & Classical & 0.45 V & $-165$&39 \\
& & GA & 0.45 V & $-165$&39 \\
\cline{2-6}
& \multirow{2}{*}{$h=3$} & Classical & 0.17 V & $-80$&39 \\
& & GA & 0.17 V & $-80$&39 \\
\cline{2-6}
& \multicolumn{5}{c}{{Effective rotation (GA): $\varphi_{\text{eff}} = 63.58^\circ$}, PF = 0.44} \\
\hline
\end{tabular}
\end{table}

Beyond numerical equivalence, the key distinction lies in the computational structure. The classical method requires $N$ independent phasor calculations followed by summation, a process that scales linearly with the number of harmonics. In contrast, the GA method computes a single operator $\bm{\Theta} = k\bm{R}$ that directly relates the total input and output waveforms. While the construction of this operator involves summing over harmonics internally (as detailed in Section III), the critical difference is that the circuit solution itself is a direct transformation  (either $\bm{i} = \bm{\Theta}_s\bm{u}$ for series circuits, or $\bm{u} = \bm{\Theta}_p\bm{i}$ for parallel circuits). It is critical to note that $\bm{\Theta}_p$ is {not} the inverse of $\bm{\Theta}_s$ (i.e., $\bm{\Theta}_p \neq \bm{\Theta}_s^{-1}$), these are two distinct operators that solve two topologically different problems. 

The proposed procedure based on GAs and real numbers has a lower computational cost than that required by the traditional phasor analysis methodology.  Moreover, the GA framework provides geometric insight that is inaccessible through classical phasors. The effective rotation angle $\varphi_{\text{eff}}$ and the bivector structure of $\bm{R}$ reveal how the circuit's impedance-like behavior manifests as a rotation in a higher-dimensional space. The power factor emerges naturally as the scalar part of this geometric transformation, eliminating the need for artificial separation into displacement and distortion factors. These interpretations are not merely conceptual conveniences—they reflect the true geometric nature of AC circuit behavior under non-sinusoidal conditions.

\subsection{Operator Analysis and Physical Interpretation}
\label{subsec:operator_analysis}

Having validated the framework's accuracy through concrete examples, the fundamental properties of the rotoflex operator are now analized and its physical interpretation is explored. This provides deeper insight into the geometric nature of circuit behavior under non-sinusoidal conditions.

\subsubsection{General Properties of the Rotance Operator}

From the theoretical development and case studies, several key properties of the rotance operator $\bm{R}$ emerge:

\begin{itemize}
    \item Unit Magnitude: The rotance always satisfies $\|\bm{R}\|=1$, as it performs a pure rotation without scaling. All scaling is captured by the flextance $k$.
    
    \item Identity for Resistive Loads: For purely resistive circuits, the voltage and current waveforms are in phase regardless of harmonic content, resulting in $\bm{R}_s = \bm{R}_p = 1$ (the scalar identity).
    
    \item Multivector Structure: For any load containing reactance, the rotance is a general multivector containing both scalar and bivector components. The scalar part represents the in-phase relationship, while the bivector parts encode the phase displacements across different harmonic planes.
    
    \item Source-Load Coupling: The specific form of $\bm{R}$ depends on \textit{both} the circuit parameters ($R, L, C$) and the harmonic content of the source. This reflects the physical reality that the overall phase relationship is determined by how each harmonic interacts with the load's frequency-dependent impedance.
\end{itemize}

\subsubsection{Geometric Interpretation of Power Factor}

The GA framework provides a direct geometric definition of the true power factor for non-sinusoidal systems. The power factor is simply the cosine of the angle between the total voltage and current vectors in the multivector space:
\begin{equation}
\text{PF} = \cos(\varphi) = \frac{\bm{u} \cdot \bm{i}}{\lVert\bm{u}\rVert \lVert\bm{i}\rVert} = \hat{\bm{u}} \cdot \hat{\bm{\imath}}
\end{equation}

This single scalar value encapsulates the combined effects of phase displacement and waveform distortion without requiring their artificial separation into displacement or distortion factor. Remarkably, this value is equivalent to the scalar part of the rotance operator: $\text{PF} = \langle \bm{R} \rangle_0$. Thus, the geometric operator that relates voltage to current {directly contains} the system's power factor as one of its components.



\subsubsection{Behavior of Canonical Loads}
\label{subsubsec:canonical_loads}

The framework's validity is further confirmed by examining its behavior for ideal $R$, $L$, and $C$ components, where it naturally reduces to well-known results:

\begin{itemize}
    \item Pure Resistance: For a purely resistive load, the rotance becomes the identity ($\bm{R}=1$) regardless of topology, and the flextances reduce to Ohm's law: $k_s = 1/R$ (series) and $k_p = R$ (parallel).
    
    \item Pure Inductance: The framework correctly reproduces the $90^\circ$ phase lag of current with respect to voltage (represented in vector GA using a counter-clockwise rotation, see \cite{montoya2021geometricteaching}). In the series case, $k_s = 1/(h\omega L)$ and $\bm{R}_s = -\bm{\sigma}_{12}$. In the parallel case, $k_p = h\omega L$ and $\bm{R}_p = \bm{\sigma}_{12}$.
    
    \item Pure Capacitance: The framework produces the complementary $90^\circ$ phase lead of current with respect to voltage (clockwise rotation). In the series case, $k_s = h\omega C$ and $\bm{R}_s = \bm{\sigma}_{12}$, while in the parallel case, $k_p = 1/(h\omega C)$ and $\bm{R}_p = -\bm{\sigma}_{12}$.
\end{itemize}

These special cases, derived naturally from the general framework, confirm its theoretical consistency with fundamental circuit principles. 

\section{Conclusions and Future Work}

This paper has introduced a complete mathematical framework based on Geometric Algebra (GA) that, for the first time, provides a direct solution method for electrical circuits under multi-harmonic, steady-state conditions in the frequency domain. The classical approach, reliant on per-harmonic decomposition and superposition, is an indirect and fragmented process ill-suited for the complexity of modern power systems. In contrast, our methodology treats non-sinusoidal waveforms as vectors in a higher-dimensional space, allowing the entire system to be solved in a single, unified mathematical operation.

The key innovation is the introduction of the {rotoflex operator} ($\bm{\Theta}$), a multivector that combines scaling and rotation into one geometric entity. Its components, the {flextance} ($k$) and the {rotance} ($\bm{R}$), are derived directly from the physical properties of the circuit and the harmonic content of the source. This provides a profound conceptual advantage, as circuit impedance is no longer just an abstract complex number but a concrete geometric transformation operator. It has been demonstrated that this general framework elegantly simplifies to classical results for canonical $R$, $L$, and $C$ loads and provides a direct geometric definition of the true power factor.

The method’s accuracy has been validated against traditional phasor analysis, and its native algorithmic implementation also proves to be computationally efficient. 
The unified formulation is inherently compatible with vector and parallel computing architectures, offering a path toward high-performance implementations that is conceptually unavailable to the sequential superposition method. This work, therefore, establishes GA not merely as an alternative, but as a conceptually and computationally superior successor to phasor analysis, opening a new and promising avenue for modern electrical engineering.

These avenues include extending the methodology to solve complex electrical networks by developing rules for combining rotoflex operators in series, parallel, and mesh configurations, thus creating a full GA-based network algebra. Another direction involves implementation and performance analysis, which requires developing optimized software libraries (e.g., in Python, Julia, or low-level implementations in C++/CUDA) and benchmarking computational performance against FFT-based superposition methods, especially for systems with a large number of harmonics. Furthermore, the framework can be extended to the analysis of electrical machines and drives, where GA's ability to handle rotations in multiple planes could represent field transformations more intuitively than traditional d-q transforms. Finally, GA's native ability to unify vector calculus and algebra can be leveraged to create multiscale models that bridge lumped-element circuit theory with Maxwell's equations for high-frequency, signal integrity, and EMC applications.

\bibliography{mybib}

\end{document}